\documentclass[12pt,english,floatfix,superscriptaddress,aps,preprint,amsmath,amssymb]{revtex4}  
\usepackage[latin9]{inputenc}
\usepackage{amsmath}
\usepackage{amssymb} 
\usepackage{amsbsy}
\usepackage{amsfonts}
\usepackage{amsopn}
\usepackage{amstext}
\usepackage{graphicx}
\usepackage{esint}
\usepackage{hyperref}

\usepackage{amssymb}
\usepackage{amsfonts}
\usepackage{amsmath}
\usepackage{graphicx}
\usepackage[english]{babel}
\usepackage{color}
\usepackage[dvips]{epsfig}
\usepackage[dvips]{graphicx}
\usepackage{float}
\usepackage{units}
\usepackage{textcomp}
\usepackage{babel}

\def \be {\begin{equation}}
\def \ee {\end{equation}}
\def \bea{\begin{eqnarray}}
\def \eea{\end{eqnarray}}
\def \ba {\begin{align}}
\def \ea {\end{align}}

\def \a {\alpha}

\def \g {\gamma}

\def \m {\mu}
\def \n {\nu}

\def \s {\sigma}

\def \o {\omega}
\def \O {\Omega}

\def \z {\zeta}

\def \p {\partial}
\def \f {\frac}

\def \nn {\nonumber}

\def \ra {\rightarrow}

\def \wt {\widetilde}

\def \inf {\infty}

\def \la {\label}

\def \be {\begin{equation}}
\def \ee {\end{equation}}
\def \ba {\begin{array}}
\def \ea {\end{array}}
\def \bea{\begin{eqnarray}}
\def \eea{\end{eqnarray}}

\def \a {\alpha}

\def \g {\gamma}

\def \m {\mu}
\def \n {\nu}

\def \s {\sigma}

\def \o {\omega}
\def \O {\Omega}

\def \z {\zeta}

\def \p {\partial}
\def \f {\frac}

\def \nn {\nonumber}

\def \ra {\rightarrow}

\def \wt {\widetilde}

\def \inf {\infty}

\def\p{\partial }

\def\beq{\begin{equation}}
\def\eeq{\end{equation}}
\def\ba{\beq\begin{array}{c}}
\def\ea{\end{array}\eeq}
\def \bea {\begin{eqnarray}}
\def \eea {\end{eqnarray}}
\def\be{\ba}
\def\ee{\ea}
\def\nn{\nonumber}
\def\p{\partial}

\def\a{{\alpha}}
\def \g {\gamma}

\def \s {\sigma}

\def \o {\omega}
\def \O {\Omega}

\begin{document}

\title{Hawking radiation of a multi-fractional Schwarzschild black hole}

\author{Peng Huang}
\email{hp@zcmu.edu.cn}
\affiliation{Department of Information, Zhejiang Chinese Medical University, Hangzhou
310013, China}

\author{Fang-Fang Yuan}
\email{ffyuan$_$hbnust@163.com}
\affiliation{Department of Physics, Hebei Normal University of Science and Technology,
Qinhuangdao 066004, China}

\vskip 1.5cm

\begin{abstract}
Following the initial work of Calcagni et al. on the black holes in multi-fractional theories,
we focus on the Schwarzschild black hole in multi-fractional theory with $q$-derivatives.
After presenting its Hawking and Hayward temperatures in detail,
we verify these results by appealing to the well-known Hamilton-Jacobi and null geodesic methods
of the tunnelling approach to Hawking radiation.
A special emphasis is placed on the difference between the geometric and fractional frames.
\end{abstract}

\maketitle

\section{Introduction}

Various theories of quantum gravity have predicted or implied the change of the effective dimension of
spacetime with the probed scale.
As a somewhat radical approach to this dimensional flow,
the multi-fractional theories consist in the direct modification of
the spacetime measure and the usual integer derivatives \cite{Calcagni:2011sz, Calcagni:2016xtk, Calcagni:2016azd}.
The construction of these multi-fractional actions are along the following four independent ways \cite{Calcagni:2016azd, Calcagni:2013yqa}:
theory $T_1$ with ordinary derivatives, theory $T_q$ with $q$-derivatives,
theory $T_v$ with weighted derivatives and theory $T_\g$ with fractional derivatives.
For a different approach, see \cite{Nottale:2008ai}.

Among the many applications,
this multi-fractional scenario has provided some insights on the relevant particle physics and
cosmology issues \cite{Calcagni:2015mxx, Calcagni:2015xcf, Calcagni:2016ofu, Calcagni:2020ads}.   
Although these investigations mostly focus on the theory $T_q$ and theory $T_v$,
one may find diverse approaches to the cosmology with fractional derivatives  
in \cite{Roberts:2009ix, Munkhammar:2010gq, Shchigolev:2010vh, Shchigolev:2013jq, Rami:2015kha, El-Nabulsi:2015szp, Giusti:2020rul, Varieschi:2020ioh, Varieschi:2020dnd, Giusti:2020kcv, Varieschi:2020hvp}.

On the other hand,
the study of black hole solutions in multi-fractional theories has only been initiated in \cite{Calcagni:2017ymp}.
The Schwarzschild black holes in theories $T_q$ and $T_v$ have been constructed there
and some interesting properties were found.
For example,
the $T_q$ solution has an additional ring singularity
while the $T_v$ solution assumes the form of a Schwarzschild-de Sitter black hole.
Notice that fractional black holes have been previously investigated in \cite{Vacaru:2010wn} from a different viewpoint.   

In this research,
we investigate the possible multi-fractional effects on the well-known tunnelling approach to Hawking radiation \cite{Parikh:1999mf, Vanzo:2011wq}.
As an elementary step,
we concentrate on the Schwarzschild black hole in multi-fractional theory with $q$-derivatives \cite{Calcagni:2017ymp}.
With the risk of overlooking many interesting works,
we just mention that over the years, both the null geodesic and Hamilton-Jacobi tunnelling methods
have been applied to all kinds of unfractional black holes with various horizons.
For the Schwarzschild black holes in particular, see \cite{Akhmedov:2006pg, Nozari:2008rc, Banerjee:2008cf, Rahman:2012id, Sakalli:2017ewb, Gomes:2018oyd, Amirfakhrian:2018ehm, Eslamzadeh:2020rbd}.

This paper is organized as follows.
In Section \ref{sec2}, we review the $T_q$-Schwarzschild black hole solution constructed in \cite{Calcagni:2017ymp}
and highlight the difference between geometric frame and fractional frame.
Based on the notations here,
we then obtain the Hawking and Hayward temperatures for this black hole in Section \ref{sec3}.
The results in fractional frame shall be corroborated by the Hamilton-Jacobi and null geodesic tunnelling methods
in Section \ref{sec4} and Section \ref{sec5}.
We also discuss the case of geometric frame in Appendix \ref{app} for completeness.

\section{Schwarzschild black hole in theory $T_q$}   \la{sec2}

As an approximation of the multi-fractional theory $T_\g$ with fractional derivatives,
the theory $T_q$ with $q$-derivatives is much easier to deal with,      
and has the merit to be invariant under Poincar\'e transformations on the geometric
coordinates \cite{Calcagni:2011sz}.
It also has the structure of a fractal spacetime
\footnote{For the difference between multiscale, multi-fractional, and multi-fractal geometries, see \cite{Calcagni:2016azd}.}.
Here we collect the basics of the $T_q$-Schwarzschild black hole constructed in \cite{Calcagni:2017ymp},
and fix the notations to facilitate our discussions in later sections.

In the multi-fractional theory $T_q$, the gravitational actions can be found by making the following substitutions:
$x^\m\ \rightarrow\ q^\m(x^\m);\ \p_\m\ \rightarrow\ v^{-1}_\m \p_\m$ with $v_\m = \p_\m q^\m$.
Correspondingly, a measure factor/profile $v(x) = \prod_\m v_\m(x^\m)$ need to be added.
As argued to be a very effective first-order coarse-graining approximation of the most general multi-fractional
measure \cite{Calcagni:2016xtk}, the following binomial measure without log oscillations has been used in \cite{Calcagni:2017ymp}
to obtain the Schwarzschild solution:
$q^\m(x^\m) = x^\m + \f{l_\ast}{\a} {\rm sgn}(x^\m) |\f{x^\m}{l_\ast}|^\a$ with $0\leq \a \leq 1$.
For the spherical coordinates, it is a useful approximation to only deform the radius as
\bea  \la{qa}
q^{(\a)} = r \Big| 1 \pm \f{1}{\a} l^{1-\a}_\ast r^{\a-1} \Big|\ .
\eea
Furthermore, the geometric radius in the theory $T_\g$ is $q(r)$ exactly.
For both theoretical and experimental reasons,
the case with $\a = \f{1}{2}$ is rather unique since it constitue a critical point for phase transition.
This value of multi-fractional index $\a$ is systematically assumed in \cite{Calcagni:2017ymp}.

In the geometric frame (integer picture) with geometric coordinate $q$,
the $T_q$-Schwarzschild black hole has the same form as the usual metric in general relativity:
\bea   \la{gm}
ds^2 &=& - f(q) dt^2 + f^{-1}(q) dq^2 + q^2 d \O^2\ ,  \\
&& f(q) = 1 - \f{r_0}{q}\ , \quad  r_0 \equiv 2GM\ .
\eea
For $\a = \f{1}{2}$, the geometric radius here is $q = | r \pm 2 \sqrt{l_\ast r} |$.
Assuming $r>4l_\ast$, we will leave out the absolute value symbol.
The locations of the horizons can be found as:
\bea   \la{hz}
\sqrt{r_h} \equiv \sqrt{r^{\pm}_h} = \sqrt{r_0 + l_\ast} \pm \sqrt{l_\ast}\ .
\eea

To make the transition  to fractional frame (fractional picture), we use the simple relation as:  
$dq = \p_r q \cdot dr \equiv v_r \cdot dr$.
Since only the spherical radius has been deformed, the measure factor reduces to $v=\prod_\m v_\m \equiv v_r$.  
To avoid the cluster of notations, we will write $v_r$ as $v$ keeping in mind that
conceptly speaking they are not the same thing.
Therefore, the $T_q$-Schwarzschild black hole in fractional frame with fractional coordinate $r$ is given by
\bea   \la{fm}
&& ds^2 = - f(r) dt^2 + f^{-1}(r) v^2 dr^2 + q^2(r) d \O^2\ ,   \\
&& f(r) \equiv f(q(r))= 1 -\f{r_0}{r \pm 2 \sqrt{l_\ast r}} \ ,\quad v \equiv v_r = 1 \pm \sqrt{\f{l_\ast}{r}} \ .
\eea

As will be clear below, almost all the discussions in this paper actually apply to the case with general multi-fractional index $\a$.
Then we would have
\bea
v^{(\a)} \equiv v^{(\a)}_r = 1 \pm l^{1-\a}_\ast r^{\a-1} \ .
\eea
Here $r>\a^{-\f{1}{1-\a}}l_\ast$ has been assumed to leave out the absolute value symbol around the geometric coordinate $q$ in (\ref{qa}).
This assumption is obviously stricter than the condition for $v^{(\a)}_r$ to be positive which is $r > l_\ast$.

\section{Black hole temperature}    \la{sec3}

In the geometric frame with geometric coordinate $q$, the temperature of the $T_q$-Schwarzschild black hole (\ref{gm}) is easily found to be:
\bea   \la{gt}
T_g &=& \f{1}{4\pi} \p_q f(q)|_{r_h}  \nn  \\
&=& \f{1}{4\pi} \cdot \f{r_0}{q^2}|_{r_h} = \f{1}{4\pi r_0} \ .
\eea
Here $q|_{r_h} = r_0$ has been used.

As stressed in \cite{Calcagni:2017ymp}, all physical observables should be computed in the fractional frame.
Correspondingly, a temperature for this black hole has been proposed there as $T_h = \f{f'(r)}{4\pi}|_{r_h}$
which has an unclear physical meaning to us.
In this work, we study instead the following two more solidly founded versions of the black hole temperature.

In the fractional frame with fractional coordinate $r$,
the $T_q$-Schwarzschild black hole (\ref{fm}) is still static
and has a timelike Killing vector $\z = \p_t$.
Thus the general formula can be employed to obtain its (Killing-)Hawking temperature:
\bea   \la{ft1}
T_h &=& \f{1}{4\pi} \sqrt{f'\Big(\f{f}{v^2}\Big)'}|_{r_h}  \nn  \\
&=& \f{1}{4\pi} \sqrt{f'\f{f'}{v^2}}|_{r_h} = \f{f'}{4\pi v}|_{r_h} = \f{1}{4\pi r_0} \ .
\eea
Along with the relation $f|_{r_h} = 0$, we also have
\bea
f'|_{r_h} \equiv \p_r f|_{r_h} = \p_q f \cdot v|_{r_h} = \f{v}{r_0} \ ,
\eea
which has been used to simplify the above calculation.
Notice that this Hawking temperature is the same as the result (\ref{gt}) in the geometric frame.

For the metric in (\ref{fm}),
one can also use the Kodama vector $\wt \z = \f{1}{v} \p_t$ to define another surface gravity \cite{Kodama:1979vn,Hayward:1997jp}. 
This leads to its (Kodama-)Hayward temperature
\footnote{The relation $\wt T_h=\f{1}{v}T_h$ here bears a resemblance to an equation found in \cite{Calcagni:2017ymp}
for the $T_v$-Schwarzschild black hole in Jordan and Einstein frames.}:
\bea   \la{ft2}
\wt T_h &=& \f{1}{4\pi} \Big(\f{f}{v^2}\Big)'|_{r_h}  \nn  \\
&=& \f{f'}{4\pi v^2}|_{r_h} = \f{1}{4\pi r_0v}|_{r_h} = \f{1}{4\pi r_0\big(1 \pm \sqrt{\f{l_\ast}{r_h}}\big)} \ .
\eea
Inserting the expression (\ref{hz}) for the horizons, the final form is
\bea
\wt T_h = \f{1}{4\pi r_0} \f{1}{1 \pm \f{1}{\sqrt{\f{r_0}{l_\ast} + 1} \pm 1}} \ .
\eea

Noting the following relation   
\bea
v = \f{r_0}{2r(1-f)} + \f{1}{2}\ , \quad v|_{r_h} = \f{r_0}{2r_h} + \f{1}{2} \ ,
\eea
we obtain an equivalent form for the Hayward temperature as
\bea
\wt T_h = \f{1}{2\pi r_0\big(1 + \f{r_0}{r_h}\big)} \ .    
\eea
For the case with general multi-fractional index $\a$, the corresponding formulas are  
\bea
v^{(\a)} = \f{\a r_0}{r(1-f)} + 1 - \a\ , \quad \wt T^{(\a)}_h = \f{1}{4\pi r_0\big(1 - \a + \f{\a r_0}{r_h}\big)} \ .
\eea

It is obvious that in the geometric frame, the Killing vector and the Kodama vector coincide with each other,
and there is no need to distinguish between the two temperatures there
\footnote{In contrast with the Hawking temperature,
the Hayward temperature is clearly more relevant to dynamical black holes
whose multi-fractional versions remain to be constructed.}.
On the other hand,
the results (\ref{ft1}) and (\ref{ft2}) in the fractional frame can be corroborated by the tunneling approach to Hawking radiation
which we will discuss in the next two sections.

\section{Hamilton-Jacobi method in fractional frame}    \la{sec4}

The Hamilton-Jacobi tunneling method \cite{Srinivasan:1998ty, Shankaranarayanan:2000qv, Angheben:2005rm}
can be straightforwardly utilized to derive the temperature (\ref{gt}) in geometric frame.
Since this is basically the same as the case of unfractional static black holes with $g_{tt}g_{rr}=-1$,
we demote the corresponding discussion to Appendix \ref{app}.
In this section,
we shall focus on the case of fractional frame.

Recalling the metric in (\ref{fm}), we have the Klein-Gordon equation as
\bea
f^{-1} \p^2_t \phi + \f{f}{v^2} \Big( \p^2_r + \f{2}{q} \p_r + \p_r \ln v \cdot \p_r \Big) \phi = \f{m^2}{\hbar^2} \phi \ ,
\eea
where the angular parts have been ignored in the s-wave approximation.
In the above expression, the part $g^{rr}(\p^2_r + \p_r \ln v \cdot \p_r)$ clearly reminds us of
the multi-fractional Laplace-Beltrami operator on Minkowski spacetime in \cite{Calcagni:2016azd}.

Using the ansatz $\phi(t,r)=\exp [ - \f{i}{\hbar} I(t,r) ]$, and retaining the leading order in $\hbar$,
we obtain the following relativistic Hamilton-Jacobi equation
\bea
(\p_t I)^2 - \f{f^2}{v^2} (\p_r I)^2 - m^2 f = 0 \ .
\eea
Corresponding to the Killing energy $\o = - \p_t I$, the solution for the action $I$ is given as
\bea
I_{\pm}(t,r) &=& - \o t \pm W(r) \ ,  \nn  \\
W(r) &=& \int dr\ \f{v}{f} \sqrt{\o^2 - m^2 f} \ .
\eea

In the near horizon approximation,
an application of the method of residues (or Feynman's $i\epsilon$ prescription) leads to
\footnote{One may also try to use the polar coordinates as in \cite{Akhmedov:2006pg} to derive this result.}
\bea
{\rm Im}\ W(r) = \f{\pi \o v}{f'}|_{r_h} \ .
\eea
A technical point here is that  
$v$ and $f$ do not have the same zeros.
Besides, we have $(\f{f}{v})'|_{r_h}=\f{1}{v}f'|_{r_h}$.
Recalling the master equation (\ref{me}), we now obtain the temperature as follows
\bea   \la{fm1}
T_h = \f{f'}{4\pi v}|_{r_h} \ .
\eea
This clearly reproduces the Hawking temperature in (\ref{ft1}).


The above discussion (together with the result in Appendix \ref{app}) confirms our previous finding
that for the $T_q$-Schwarzschild black hole,
the Hawking temperature has the same expression in the geometric and fractional frames.
In the framework of the Hamilton-Jacobi tunnelling method,
it has also been proposed in \cite{Akhmedov:2006pg} that
any coordinate transformation involving
only spatial coordinates should yield the same result for the temperature.

A different perspective has been advocated in \cite{Angheben:2005rm}
which argues for the importance of proper distance to arrive at the rigorous result.
Due to the singular nature of the metric (\ref{fm}),
the proper radial distance here is not $r$, but  
\bea
\s = \int dr\ \f{v}{\sqrt{f}} \ .  
\eea
In this way, assuming the near horizon expansion, we would have
\bea
W(\s) = \f{2v}{f'}|_{r_h} \int \f{d\s}{\s} \sqrt{\o^2 - \f{\s^2}{4} \f{f'^2}{v^2}|_{r_h} m^2} \ .
\eea
This leads to ${\rm Im}\ W(\s) = 2\ {\rm Im}\ W(r)$.
However, now we should use the master equation in regular coordinates which is
\bea   \la{me2}
-\ {\rm Im}\ I_- = \f{\o}{2T_H} \ .
\eea
Considering that the temporal contribution is zero in this case,
we again get the same result as in (\ref{fm1}).

Notice that although here $r_h$ is two-valued,
we do not need to resort to the technique in \cite{Angheben:2005rm} designed for the black holes with multiple horizons.

If using Kodama energy $\wt \o = \f{1}{v} \o$ instead, then we would have $\p_t I = - v \wt \o$.
This leads to the following solution
\bea
I_\pm = v ( - \wt \o t + \wt W) \ .
\eea
Along a similar procedure, we can reobtain the Hayward temperature as (cf. (\ref{ft2}))
\bea
\wt T_h = \f{f'}{4\pi v^2}|_{r_h} \ .
\eea

\section{Null geodesic method}    \la{sec5}

In spite of some conceptual issues \cite{Kerner:2006vu, Vanzo:2011wq},
the null geodesic tunnelling method has captured some essential features of the Hawking radiation,
and can still be used to derive the black hole temperature.
For recent applications, see \cite{Johnson:2018gjr, Amirfakhrian:2018ehm, Johnson:2019kda, Eslamzadeh:2020rbd}.

This method requires a transition to the regular Painlev\'{e}-Gullstrand coordinates
which can be implemented by the following transformation
\bea
t_p = t \pm \int dr\ \f{1}{f} \sqrt{v^2-f} \ .
\eea
Now the metric for $T_q$-Schwarzschild black hole in fractional frame becomes
\bea
ds^2 = - f(r) dt^2_p \pm 2\sqrt{v^2-f} dt_p dr + dr^2 + q^2(r) d \O^2 \ .
\eea

For an s-wave, the null geodesic is given by
\bea
\dot r \equiv \f{dr}{dt_p} = \pm v - \sqrt{v^2-f} \ .
\eea
Corresponding to an outgoing positive energy particle which crosses the horizon,
the action can be written as
\bea
I_- = \int^{r_+}_{r_-} dr\ p^{(+)}_r = - \o \int^{r_+}_{r_-} \f{dr}{\dot r_{(+)}} \ .      
\eea
Here the Hamilton's equation $\dot{r} = \f{dH}{dp_r}$ and the relation $dH=-d\o$ have been used.    

Combining the above two equations, we obviously have
\bea
I_- = \o \int^{r_+}_{r_-} dr\ \f{v}{f} \bigg( 1 + \sqrt{1 - \f{f}{v^2}} \bigg) \ .
\eea
A standard calculation then leads to
\bea
I_- = \o \cdot \pi i \cdot \f{v}{f'} \cdot 2 \ .
\eea
Recalling the master equation (\ref{me2}) in regular coordinates,
one easily obtains the Hawking temperature in fractional frame as $T_h = \f{f'}{4\pi v}|_{r_h}$(cf. (\ref{ft1})).

When setting $v \ra 1,\ r \ra q$,
the above discussion reduces to the case of geometric frame.
However, since both $v$ and $q$ depend on the fractional coordinate $r$ explicitly,
this only happens when $r \ra \inf$.
In general situations,
the multi-fractional effects may still exist in the fractional frame.


\section{Conclusion}

To sum up the main results in this work,
we firstly gave the Hawking and Hayward temperatures (\ref{ft1}, \ref{ft2}) in some detail
for the Schwarzschild black hole in multi-fractional theory $T_q$ with $q$-derivatives \cite{Calcagni:2017ymp}.
Then the Hamilton-Jacobi and null geodesic tunnelling methods were applied to rederive these results.
Although our initial motivation was to study the multi-fractional effects on the tunnelling approach itself,
at least in the case of theory $T_q$, we haven't find any notable modification.
One may attribute this to the special nature of the theory with $q$-derivatives,
and expect the situation to be different for the theory $T_\g$ with fractional derivatives.

Other possible directions for further research are as follows.
In the first place,
one may reconsider some issues which have been studied in the framework of tunnelling approach
such as fermion tunneling, charged and rotating black holes, dynamical horizons, modified gravities, etc.
Secondly,
one may trace back to the origins to revisit the Hawking radiation and Unruh effect in multi-fractional theories.
As for the black hole thermodynamics,
one may wonder if there is a possible connection to Barrow's fractal entropy \cite{Barrow:2020tzx}.
This may still necessitate the need to go beyond the multi-fractional theory $T_q$  
\footnote{For the Hawking temperature of $T_q$-Schwarzschild black hole,
the area law continues to hold as one can easily seen by use of the relation $S=\int\f{dM}{T(M)}$.}.
Finally,
besides the requisite constructions of more general multi-fractional black holes,   
their spacetime structures including the near horizon and asymptotic symmetries
may be rather interesting to investigate in depth.

\appendix

\section{Hamilton-Jacobi method in geometric frame}   \la{app}

In contrast with the original null geodesic method \cite{Parikh:1999mf} of the tunneling approach to Hawking radiation,
the Hamilton-Jacobi method \cite{Srinivasan:1998ty, Shankaranarayanan:2000qv, Angheben:2005rm}
is proved to be more transparent and rigorous and thus more widely used
\footnote{However, an argument can be given to verify the equivalence of a generalised null geodesic method
with the latter \cite{Vanzo:2011wq}.}.
Here we demonstrate this procedure to obtain the temperature (\ref{gt}) in geometric frame
\footnote{Although this Schwarzschild-like metric is in a singular gauge,
the method is still applicable.}.
The main intention is to benefit the comparison with the case of fractional frame in Section \ref{sec4}.

The starting point is the following Klein-Gordon equation
\bea
\f{1}{\sqrt{-g}}\p_\m ( \sqrt{-g} g^{\m\n} \p_\n \phi ) = \f{m^2}{\hbar^2} \phi \ .
\eea
In the s-wave approximation, the angular parts can be ignored.
For the $T_q$-Schwarzschild black hole (\ref{gm}) in geometric frame,
the above equation becomes
\bea
f^{-1}(q) \p^2_t \phi + f(q) \Big( \p^2_q + \f{2}{q} \p_q \Big) \phi = \f{m^2}{\hbar^2} \phi \ .
\eea
Using the ansatz $\phi(t,q)=\exp [ - \f{i}{\hbar} I(t,q) ]$ and invoking the WKB approximation,
to leading orger in $\hbar$, we have
\bea
(\p_t I)^2 - f^2(q) (\p_q I)^2 - m^2 f(q) = 0 \ .
\eea
This is essentially the relativistic Hamilton-Jacobi equation $g^{\m\n}\p_m I \p_\n I + m^2 = 0$.  

For the Killing energy $\o = - \p_t I$, the solution can be found to be
\bea
I_{\pm}(t,q) &=& - \o t \pm W(q) \ ,  \nn  \\
W(q) &=& \int dq\ \f{1}{f(q)} \sqrt{\o^2 - m^2 f(q)} \ .
\eea
Assuming the near horizon expansion, one has
\bea
W(q) = \int dq\ \f{1}{\p_q f(q) (q - q|_{r_h})} \sqrt{\o^2 - m^2 \p_q f(q) (q - q|_{r_h})}\ .
\eea
Notice that here $q|_{r_h}$ is actually single-valued.
At this point, we may use the method of residues (basically Feynman's $i\epsilon$ prescription) to arrive at the following result
\bea
{\rm Im}\ W(q) = \f{\pi \o}{\p_q f(q)} \ .
\eea

To obtain the black hole temperature, we need a master equation which is
\footnote{This may be interpreted e.g. as a consequence
of the principle of detailed balance $P_- = \exp (- \f{\o}{T}) P_+$ \cite{Banerjee:2008cf}.}
\bea   \la{me}
{\rm Im}\ ( I_+ - I_- ) = \f{\o}{2T_H} \ .
\eea
Notice that the covariant formalism of \cite{Vanzo:2011wq} is based on an integral form of this equation.
Additionally, for black holes in regular (such as Painlev\'{e}-Gullstrand) coordinates,
the first term on the left hand side is actually zero
since the plus sign corresponds to the ingoing particles.
Finally, we have
\bea
T_g = \f{\o}{4\ {\rm Im}\ W(q)} = \f{\p_q f(q)}{4\pi}|_{r_h} \ .
\eea
Thus the temperature (\ref{gt}) in geometric frame has been reproduced.


\end{document}